\documentclass{article}

\usepackage{spconf}
\usepackage{cite}
\usepackage{amsmath,amssymb,amsfonts}
\usepackage{algorithmic}
\usepackage{graphicx}
\usepackage{textcomp}
\usepackage{xcolor}
\usepackage[acronym]{glossaries}
\usepackage{comment}
\usepackage{array}
\usepackage{makecell}
\usepackage{tabularx}
\usepackage{tabularray}
\usepackage{siunitx}
\usepackage{booktabs}
\newcolumntype{L}{>{\raggedright\arraybackslash}X}
\usepackage{relsize}\usepackage{subdepth}\allowdisplaybreaks
\usepackage{caption}
\usepackage[labelformat=simple]{subcaption}
\usepackage{stfloats} 
\usepackage[linesnumbered,ruled,vlined]{algorithm2e}
\usepackage{array}
\usepackage{xcolor}
\SetKwInput{KwInput}{Input}                
\SetKwInput{KwOutput}{Output}              

\newacronym{tdth}{TD-TH}{time domain thresholding}
\newacronym{tfdth}{TFD-TH}{time-frequency domain thresholding}
\newacronym{fmcw}{FMCW}{frequency modulated continuous wave}
\newacronym{pd}{PD}{probability of detection}
\newacronym{sinr}{SINR}{signal-to-interference-plus-noise ratio}
\newacronym{pri}{PRI}{pulse repetition interval}
\newacronym{adc}{ADC}{analog-to-digital converter}
\newacronym{fft}{FFT}{fast Fourier transform}
\newacronym{lrr}{LRR}{long-range radar}
\newacronym{srr}{SRR}{short-range radar}
\newacronym{cfar}{CFAR}{constant false alarm rate}
\newacronym{rfe}{RFE}{radar front end}
\newacronym{lpf}{LPF}{low-pass filter}
\newacronym{rd}{RD}{range-Doppler}
\newacronym{cdf}{CDF}{cumulative distribution function}
\newacronym{adas}{ADAS}{advanced driver assistance system}
\newacronym{ad}{AD}{automonous driving}
\newacronym{stft}{STFT}{short-time Fourier transform}

\title{Performance Evaluation and Analysis of Thresholding-based Interference Mitigation for Automotive Radar Systems
}

\name{Jun Li$^{\star}$, Jihwan Youn$^{\dagger}$, Ryan Wu$^{\star}$, Jeroen Overdevest$^{\dagger}$, Shunqiao Sun$^{\ddagger}$}
\address{$^{\star}$NXP Semiconductors, San Jose, CA, USA \\
         $^{\dagger}$NXP Semiconductors, Eindhoven, The Netherlands \\
         $^{\ddagger}$ The University of Alabama, Tuscaloosa, AL, USA}
  
\begin{document}
\ninept

\maketitle

\begin{abstract}
In automotive radar, time-domain thresholding (TD-TH) and time-frequency domain thresholding (TFD-TH) are crucial techniques underpinning numerous interference mitigation methods. Despite their importance, comprehensive evaluations of these methods in dense traffic scenarios with different types of interference are limited. In this study, we segment automotive radar interference into three distinct categories. Utilizing the in-house traffic scenario and automotive radar simulator, we evaluate interference mitigation methods across multiple metrics: probability of detection, signal-to-interference-plus-noise ratio, and phase error involving hundreds of targets and dozens of interfering radars. The numerical results highlight that TFD-TH is more effective than TD-TH, particularly as the density and signal correlation of interfering radars escalate.
\end{abstract}


\section{Introduction}
Radar-to-radar interference refers to undesired signals coming from surrounding radars, which can deteriorate the performance of a victim radar by corrupting desired target signals. Several decades ago, when automotive radars were first available and utilized in high-end vehicles, radar interference has not been considered as a significant issue given their limited deployment. Consequently, interference mitigation algorithms were mainly confined to basic time domain detection and processing. Today, automotive radars are ubiquitous and play an instrumental role in various \gls{adas} applications such as adaptive cruise control, autonomous emergency brake, blind spot detection, line change assistance, etc \cite{SUN_SPM_Feature_Article_2020,sun20214d}. The cutting-edge 4D automotive radar has further cemented its importance in the domain of autonomous driving (AD) owing to its high-resolution capability and resilience against adverse weather conditions \cite{Igal_Bilik2019,blrc2023}. As the use of automotive radars proliferates, radar-to-radar interference is inevitable and intensifying, underscoring the pressing need for innovative interference mitigation strategies \cite{Alland_Interference_SPM_2019}.

In recent years, numerous automotive radar interference mitigation methods \cite{bechter2017automotive,fischer2016untersuchungen,marvasti2012sparse,tullsson1997topics,toth2019performance,JeroenTDAT,wang2021cfar,wu2022radar,wu2023radar,tanis2018automotive,muja2022interference} have been proposed by the academic and industrial communities. Central to these methods are two core techniques: \gls{tdth} and \gls{tfdth}. While many advanced interference mitigation methods build upon these two techniques by incorporating additional steps after thresholding, such as interpolation to enhance signal-to-noise ratio \cite{toth2019performance,wang2021cfar,marvasti2012sparse}, the crux of their performance lies in the efficacy of \gls{tdth} and \gls{tfdth}. The importance of \gls{tdth} and \gls{tfdth} is not merely due to their role as bases for advanced interference mitigation methods; it is also attributed to their inherent applicability. As interference mitigation typically precedes other signal processing steps (e.g., range, velocity, and angle estimation) in automotive radar processors, and given the real-time requirements of such systems, both techniques stand out as essential and pragmatic with their lightweight computation. However, a comprehensive comparison between them in dense traffic scenarios with different types of interference remains absent. This gap in analysis leaves the potential and advantages of time-frequency domain processing not fully explored or understood.

In this paper, we aim to bridge the gap by comparing the performance of \gls{tdth} and \gls{tfdth} in a more comprehensive setup compared to the prior studies. It is imperative to understand that conducting real-world experiments with dozens of moving interference sources is challenging or virtually impossible given the safety concerns involved. Thus, hundreds of targets (e.g., vehicles and guardrails) and dozens of interfering radars with various types of interference are simulated by our in-house traffic scenario simulator and high-fidelity automotive \gls{rfe} simulator. Then, the interference mitigation performance of \gls{tdth} and \gls{tfdth} was evaluated in terms of \gls{pd} and \gls{sinr} in the \gls{rd} map level, unlike other works that did the evaluation in the range spectrum level. Furthermore, we introduce phase error \gls{cdf} as a significant metric to evaluate the influence of interference mitigation on subsequent radar signal processing, such as angle estimation, which has been overlooked in most automotive radar interference mitigation works.

\section{System Model}
\begin{figure}
\centering
  \includegraphics[width=0.4\textwidth]{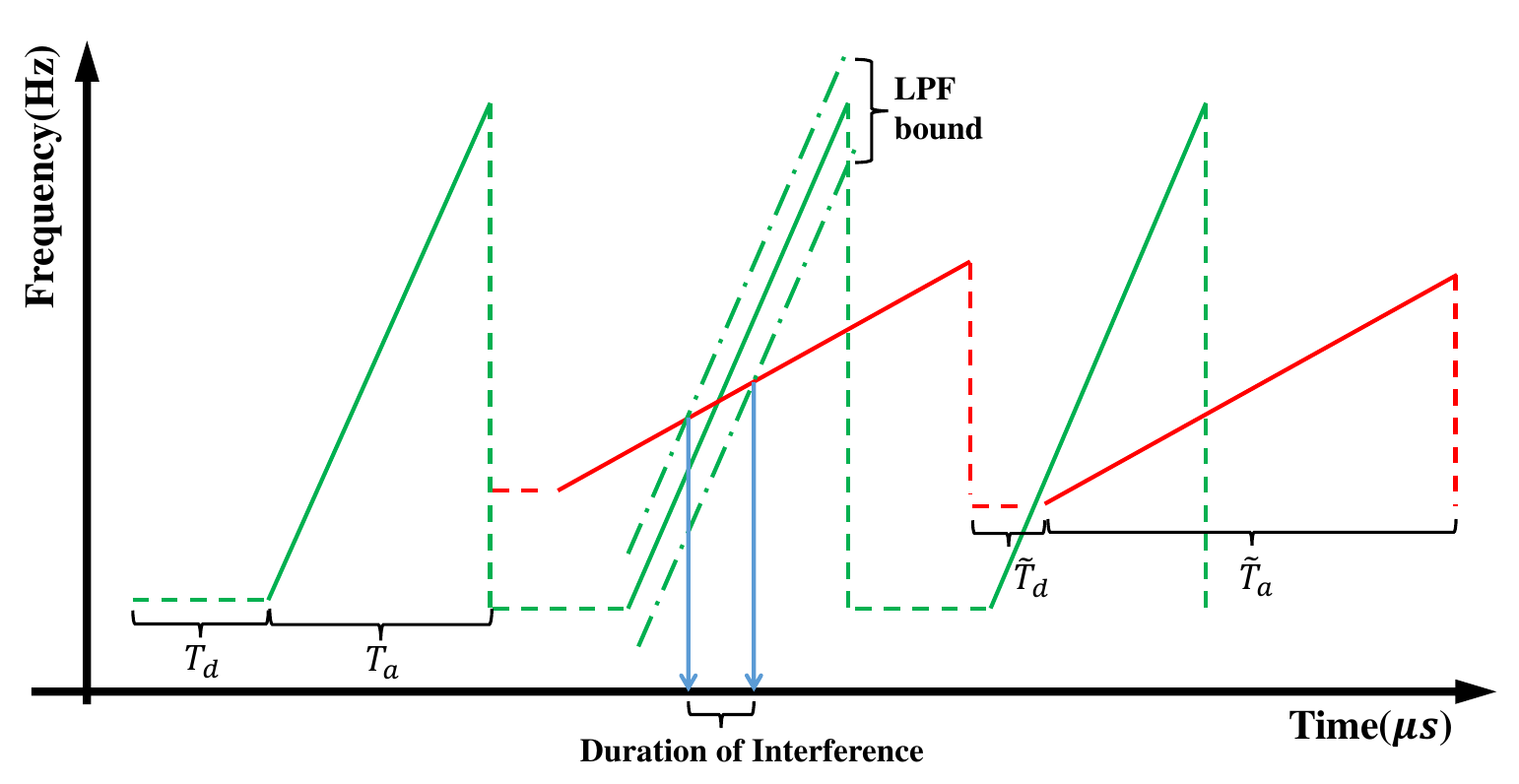}
  \caption{Duration of interference overview. Green chirps are signals transmitted by a victim radar and red chirps are interference signals transmitted by an interfering radar.}
  \label{fig:waveform}
\end{figure}

Consider a simple traffic scenario comprising a victim radar, an interfering radar, and a single point target. The transmitted \gls{fmcw} signal of the victim radar is given by:
\begin{equation}
    s_p(t)=e^{j\pi\alpha t^2},\,\, 0<t<T_a,
    \label{eq:sp}
\end{equation}
where $\alpha$ is the chirp slope determined by $\alpha=\frac{B}{T_a}$. Here, $T_a$ indicates the active time duration of a single pulse, and $B$ is the chirp sweep bandwidth. Given that $M$ pulses are transmitted per coherent processing interval, the transmitted modulated signal of the victim radar can be expressed as
\begin{equation}
    s_\text{tx}(t)=\sum_{m=0}^{M-1}s_p(t-mT_\text{PRI})e^{j2\pi f_ct},
    \label{eq:stx}
\end{equation}
where $f_c$ is the carrier frequency, and $T_\text{PRI}$ is the \gls{pri}, combining the active and dead time ($T_\text{PRI}=T_a+T_{d}$ and $T_{d}>0$), as shown in Fig.~\ref{fig:waveform}. The interfering radar emits a signal, $\tilde{s}_\text{tx}(t)$, resembling the waveforms defined by (\ref{eq:sp}) and (\ref{eq:stx}), but with varied parameters. In this paper, a tilde notation will be used for interference to differentiate these parameters from those of the victim radar.

\subsection{Target Signal Model}
The transmitted signal reflected back by a single target is received and demodulated at the \gls{rfe}. The received signal, after passing through the mixer, can be represented by
\begin{equation}
    \begin{split}
        s_\text{rx}(t)&=as_\text{tx}(t-\tau)\, s_\text{tx}^*(t)\\
    \end{split}
\end{equation}
where $a$ indicates the complex target amplitude and $\tau$ is the time delay due to the round-trip propagation between the victim radar and the target, which can be approximated by
\begin{math}
    \tau\approx\frac{2(r+\dot{r}mT_\text{PRI})}{c},
\end{math}
where $c$ is the speed of light, and $r$ and $\dot{r}$ represent the radial distance and velocity of the target, respectively. 

After the received signal is filtered by a \gls{lpf} and sampled by an \gls{adc} with a time interval $T_s$, the \gls{adc} output can be framed in terms of slow time $m$ and fast time $n$ (i.e., $t=mT_\text{PRI}+nT_s$) with $N$ active samples per \gls{pri}, $n\in\{0,1,2,...,N-1\}$:
\begin{equation}
    s_\text{rx}(m,n)=ae^{-j2\pi f_c\frac{2r}{c}}e^{-j2\pi f_rn}e^{-j2\pi f_dm},
    \label{eq:smn_tgt}
\end{equation}
where $f_r=\frac{2r\alpha T_s}{c}$ denotes the normalized range frequency, and $f_d=\frac{2\dot{r}T_\text{PRI}}{\lambda}$ represents the normalized Doppler frequency \cite{jin2022spatial}. By applying the \gls{fft} along the $n$-dimension (i.e., range \gls{fft}) and the $m$-dimension (i.e., Doppler \gls{fft}), one can extract information on range and velocity. However, under the presence of interference, the target information can be obscured after the range and Doppler \gls{fft} processing.

\subsection{Interference Signal Model}
The demodulated and dechirped interference signal at the receiver can be expressed as 
\begin{equation}
    \begin{split}
        \tilde{s}_\text{rx}(t)=\tilde{a}\tilde{s}_\text{tx}(t)\, s_\text{tx}^*(t),
        \label{eq:stx_intf1}
    \end{split}
\end{equation}
where $\tilde{a}$ indicates the complex interference amplitude. It is worth noting that representing (\ref{eq:stx_intf1}) analytically is challenging due to unknown and varied configurations of the interfering radar. Without loss of generality, we focus on the time period at $mT_\text{PRI}\leq t\leq mT_\text{PRI}+T_a$ with $\tilde{f}_c=f_c$, which is the worst interference scenario. Assuming the $\tilde{m}^\text{th}$ interference chirp is mixed with the reference signal in this time period, the received interference signal can be obtained as 
\begin{equation}
    \begin{split}
        \tilde{s}_\text{rx}(m,n)=\tilde{a}e^{j\pi(\tilde{\alpha}-\alpha)(nT_s)^2}g(n,m,\tilde{m},\tilde{\alpha},\alpha,T_\text{PRI},\tilde{T}_\text{PRI})
        \label{eq:stx_intf2}
    \end{split}
\end{equation}
From (\ref{eq:stx_intf2}), we can interpret the interference signal obtained at the \gls{adc} output as a type of chirp signal, termed here as the post-mix chirp. Its chirp slope is equal to the chirp slope difference between the victim radar signal and the interfering radar signal, i.e., $\tilde{\alpha}-\alpha$. Furthermore, we utilize a generic function, $g(\cdot)$, to denote that the interference's starting frequency and initiation time are functions of $\tilde{\alpha},\alpha, \tilde{T}_\text{PRI}$, and $T_\text{PRI}$, with each chirp being unique. The duration of interference $\delta$ depends on both the post-mix chirp slope and the \gls{lpf} of the victim radar \cite{uysal2018mitigation}:
\begin{equation}
    \begin{split}
        \delta = \frac{1}{T_s|\tilde{\alpha}-\alpha|}
        \label{eq:delta}
    \end{split}
\end{equation}

\subsection{Characteristic of Received Signal} \label{subsec:intf}

The received signal $y(m,n)$ including target signal, interference signal, and white Gaussian noise can be obtained from (\ref{eq:smn_tgt}) and (\ref{eq:stx_intf2}), and written in a matrix format as
\begin{equation}
    \begin{split}
        \mathbf{Y}=\mathbf{S}_\text{rx}+\tilde{\mathbf{S}}_\text{rx}+\mathbf{E},
    \end{split}
\end{equation}
Here, $\mathbf{Y},\mathbf{S}_\text{rx}, \tilde{\mathbf{S}}_\text{rx}$, and $\mathbf{E}$ are all of dimension $M\times N$. Every entry, denoted as $(i,j)$, within these matrices corresponds to $y(i,j),{s}_\text{rx}(m,n),\tilde{s}_\text{rx}(m,n)$, and $\epsilon(m,n)$ respectively. Within this framework, $i$ is the chirp index, and $j$ represents the $j^\text{th}$ \gls{adc} samples in the $i^\text{th}$ chirp with $i\in\{1,...,M\}$ and $j\in\{1,...,N\}$. The term $\epsilon(m,n)$ represents the white Gaussian noise in our model.  

\begin{figure*}[hbt!]
\centering
\includegraphics[width=6.5 in]{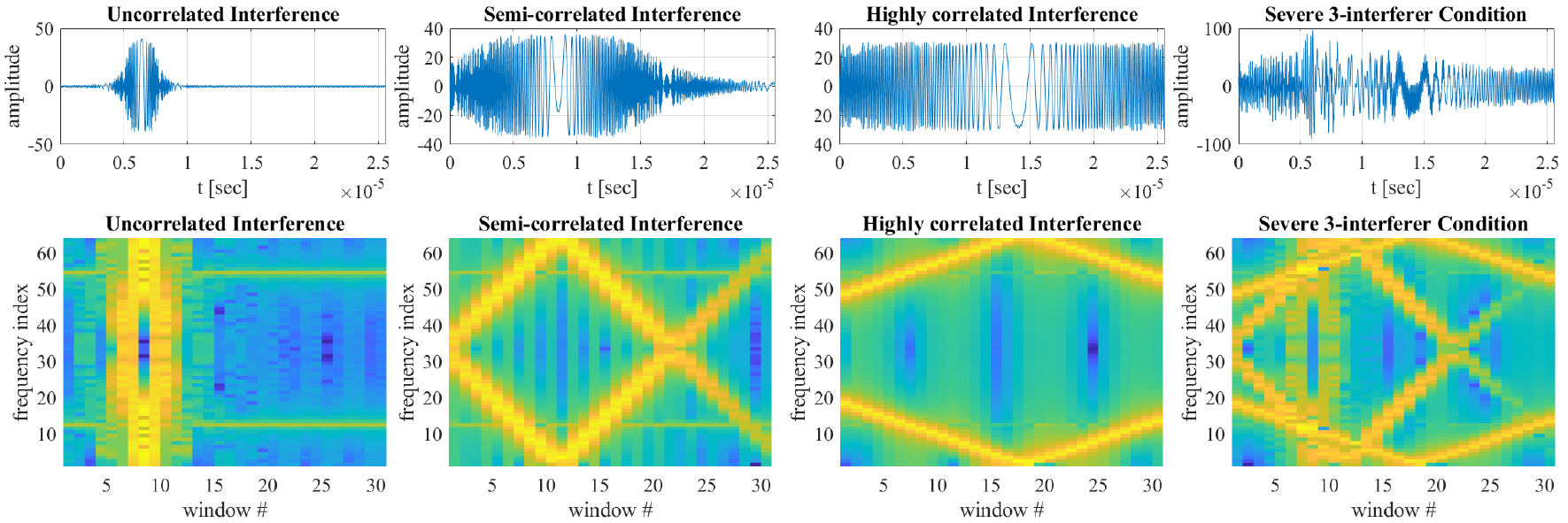}
\caption{Different types of interference with a single target in time domain (top) and time-frequency domain (bottom). From left to right, uncorrelated ($\gamma=0.5$), semi-correlated ($\gamma=1.5$), highly correlated interference ($\gamma=9$), and a case interfered by three types of interference.}
\label{fig:intf_types}
\end{figure*}

In modern automotive radar, there are various waveform configurations for different radar applications \cite{terbas2019radar}. Based on different $\tilde{\alpha}$, we can classify interference signals into 3 categories using the chirp decorrelation factor $\gamma$, which follows a distribution $p(\gamma)$ with $\gamma\in (10\%,1000\%)$. The interference signal chirp slope can be associated with the victim radar chirp slope by $\gamma$ as
\begin{equation}
    \begin{split}
        \tilde{\alpha}=\alpha\pm\frac{f_s}{N_s/f_s}\gamma.
        \label{eq:gamma}
    \end{split}
\end{equation}
Then, interference can be characterized as:
\begin{itemize}
    \item \textbf{Uncorrelated Interference:} Identified by a post-mix chirp that scans through a range of $(200\%, 1000\%)\times f_S$ within $T_a$. As illustrated in Fig.~\ref{fig:intf_types}, in the time domain, the target signal is a sinusoidal signal, while the interference signal can be considered as transient impulse noise. In the time-frequency domain, the interference signal presents as low-pass filtered linear-ramp signal, i.e., post-mix chirp, as opposed to target signal seen as a flat tone. 
    \item \textbf{Semi-correlated Interference:} Defined by a post-mix chirp with a slope that scans through $(75\%, 200\%)\times f_S$ within $T_a$. According to (\ref{eq:delta}) and (\ref{eq:gamma}), the smaller the $\gamma$ is, the smaller the post-mix chirp slope is, leading to a longer duration of interference $\delta$ within $T_a$. 
    \item \textbf{Highly correlated Interference:} Characterized by a post-mix chirp with a slope that scans through $(10\%, 75\%)\times f_S$ within $T_a$. Significantly, for this type of interference, the interference duration extends to envelop the entire chirp time. Consequently, all time samples are tainted by interference, as shown in Fig.~\ref{fig:intf_types}.
\end{itemize}

\section{Thresholding-based Interference Mitigation Algorithms} \label{sec:th}
In this section, we introduce the core techniques of various thresholding based interference mitigation methods, namely \gls{tdth} and \gls{tfdth}. While \gls{tdth} methods operate directly in the time domain to detect and eliminate interference, \gls{tfdth} methods work in the time-frequency domain for the same purpose. A variety of time-frequency representations \cite{boashash2015time,selesnick2009short} can be used, alongside various interference detectors such as \gls{cfar}\cite{JeroenTDAT,wang2021cfar} or L-statistics\cite{wu2022radar,muja2022interference}, to establish the threshold value $\beta$ for determining interference. However, the fundamental steps for thresholding are consistently applied across methods, as detailed in Algorithm~\ref{alg:TD} and \ref{alg:TFD}. Note that the 
\gls{stft} is presented here merely as an example. An in-depth discussion on the advantages and disadvantages of different time-frequency representations exceeds the scope of this paper.

\SetKwComment{Comment}{/* }{ */}
\RestyleAlgo{ruled}
\begin{algorithm}[ht!]
\DontPrintSemicolon
  \SetAlgoLined
\caption{Time Domain Thresholding (TD-TH)}\label{alg:TD}
\KwInput{Received \gls{adc} Samples $\mathbf{Y}$}
\KwOutput{$\mathbf{Y}_c$ after interference mitigation}
    $[M,N]=\mathbf{size}\{\mathbf{Y}\}$\;
    \For{$i=1:M$ }
        {$\mathbf{x} = \mathbf{Y}(i,:)$\;
        $\beta = \mathbf{IntfDetector}\{\mathbf{x}\}$\;
        $\mathbf{x}(\mathbf{x}>\beta) = 0$\;
$ \mathbf{Y}_c(i,:) = \mathbf{x}$\;
}
\end{algorithm}

\SetKwComment{Comment}{/* }{ */}
\RestyleAlgo{ruled}
\begin{algorithm}[ht!]
\DontPrintSemicolon
  \SetAlgoLined
\caption{Time-Frequency Domain Thresholding (TFD-TH)}\label{alg:TFD}
\KwInput{Received \gls{adc} Samples $\mathbf{Y}$}
\KwOutput{$\mathbf{Y}_c$ after interference mitigation}
    $[M,N]=\mathbf{size}\{\mathbf{Y}\}$\;
    \For{$i = 1:M$ }
        {$\mathbf{x} = \mathbf{Y}(i,:)$\;
        $\mathbf{X} = \mathbf{STFT}\{\mathbf{x}\}$\;
        $[N_r,N_c]=\mathbf{size}\{\mathbf{X}\}$\;
        
        \For{$k=1:N_r$}
        {
        $\mathbf{z}=\mathbf{X}(k,:)$\;
        $\beta =\mathbf{IntfDetector}\{\mathbf{z}\}$\;
        $\mathbf{z}(\mathbf{z}>\beta)=0$\;
        $\mathbf{X}_c(k,:)=\mathbf{z}$\;
        }
        $\mathbf{Y}_c(i,:) = \mathbf{STFT}^{-1}\{\mathbf{X}_c\}$\;
}
\end{algorithm}
Comparing Algorithm~\ref{alg:TD} with Algorithm~\ref{alg:TFD}, it is evident that \gls{tdth} methods typically have lower computational complexity than \gls{tfdth} methods, primarily due to the additional STFT and inverse STFT operations in \gls{tfdth}. 

From (\ref{eq:delta}) and Fig.~\ref{fig:intf_types}, it is clear that in the presence of uncorrelated interference, only a minority of the \gls{adc} time samples are tainted by interference. This reduced impact is a direct result of the deramp mixing and the \gls{lpf} in the receive chain. Therefore, after applying \gls{tdth}, we can still employ the uncorrupted time samples for target range estimation. However, as the correlation between the interfering radar signal and the victim radar signal increases, more \gls{adc} time samples are corrupted according to (\ref{eq:delta}) and Fig.~\ref{fig:intf_types}. Specifically, in instances of highly correlated interference, the chirp slope of interference closely aligns with that of the victim radar. Consequently, the interference cannot be filtered by the \gls{lpf} and the victim radar continually captures the interfering radar's signal at its \gls{adc} output. This scenario compromises the efficacy of \gls{tdth}. Furthermore, with increasing amounts of interference, another challenge emerges for \gls{tdth} as shown in the severe 3-interferer condition presented in Fig.~\ref{fig:intf_types}. Even if all interferences are uncorrelated interferences in the multiple-interferer example, their combined impact could corrupt the whole time samples. With each interference source corrupting a fraction of the \gls{adc} time samples, a point might be reached where all the samples are tainted, leaving none uncorrupted. This saturation further diminishes the performance of \gls{tdth}. However, different interferences appear as distinct nonzero-slope linear features in the time-frequency domain, crossing the flat tones of target signals at distinct time and frequency components. Thus, in the time-frequency domain, the interference signals are distinctly recognizable and can be effectively isolated, ensuring that the target signals remain largely unaffected. Given the scenarios described, it becomes evident that the \gls{tfdth} methods theoretically hold an advantage over the \gls{tdth} approaches.

\section{Numerical Results}
Using the previously developed traffic scenario and automotive radar \gls{rfe} simulator, radar signals with interference are generated to evaluate various interference mitigation algorithms. In this paper, we focused on two highway traffic scenarios with the targets being randomly positioned vehicles on the highway and guardrails along the edges of the highway. It is important to note that any vehicle target can potentially serve as a source of interference. The waveform configurations of interfering radars are randomly selected. Specific details are provided in Table~\ref{tab:traffic}, where Interference A, B, and C correspond to uncorrelated, semi-correlated, and highly correlated interference, as elaborated in section~\ref{subsec:intf}. In Scenario 1 (S1), uncorrelated interference predominates, accounting for $90\%$. In Scenario 2 (S2), on the other hand, highly correlated interference predominates. 

\begin{table}
\centering
\caption{\textsc{Highway Traffic Scenario}}\label{tab:traffic}
\begin{tabularx}{\linewidth}{c c c}
\toprule
Traffic setting  & Scenario 1 (S1) & Scenario 2 (S2) \\
\midrule
Number of lanes & 6 & 6   \\
Type of targets & (vehicle, guardrail) & (vehicle, guardrail)   \\
Number of targets & (34, 74) & (34, 74)   \\
Interference A & 90\% & 5\%\\
Interference B & 5\% & 5\%\\
Interference C & 5\% & 90\%\\
\bottomrule
\end{tabularx}
\end{table}

\begin{figure*}
\centering
\begin{minipage}[c]{0.3\linewidth}
\centering
\includegraphics[width=1\linewidth]{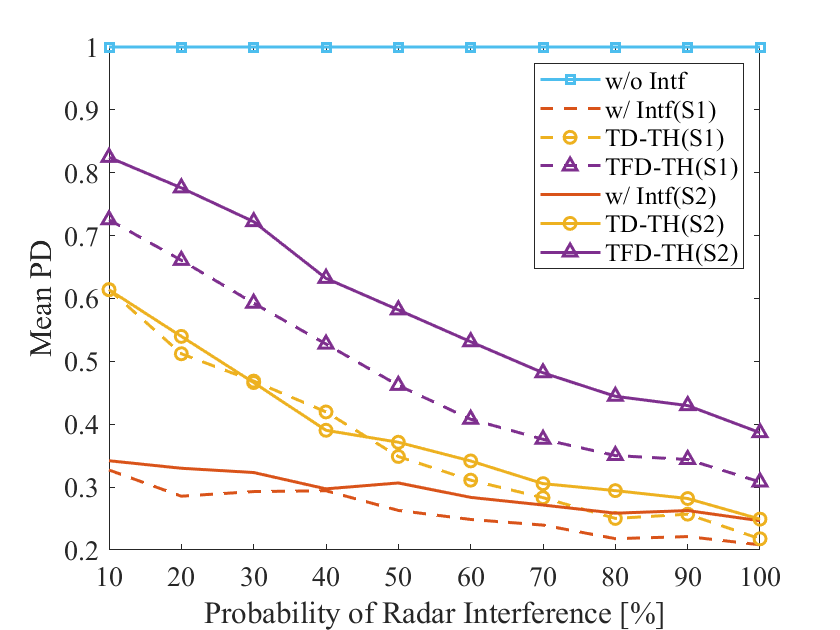}
\centerline{(a)}\medskip
\end{minipage}
\begin{minipage}[c]{0.3\linewidth}
\centering
\includegraphics[width=1\linewidth]{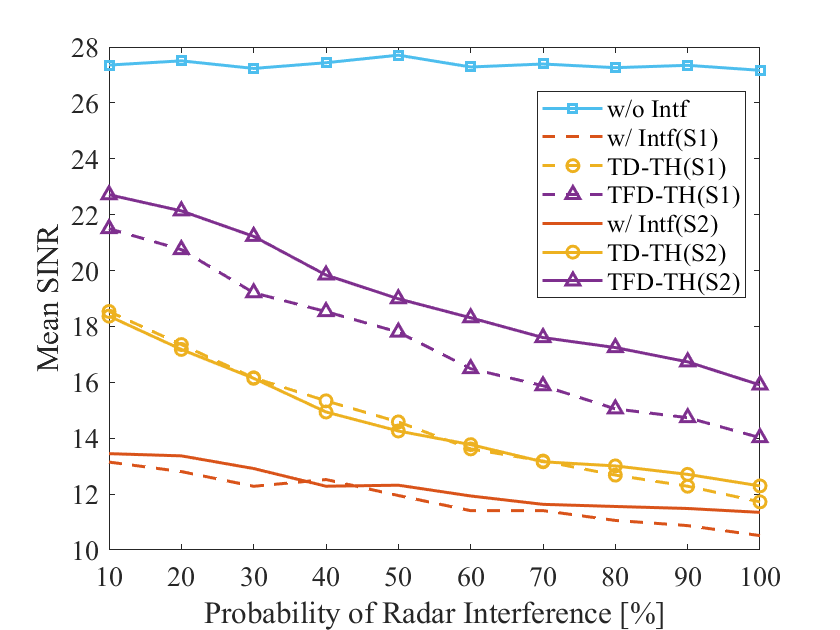}
\centerline{(b)}\medskip
\end{minipage}
\begin{minipage}[c]{0.3\linewidth}
\centering
\includegraphics[width=1\linewidth]{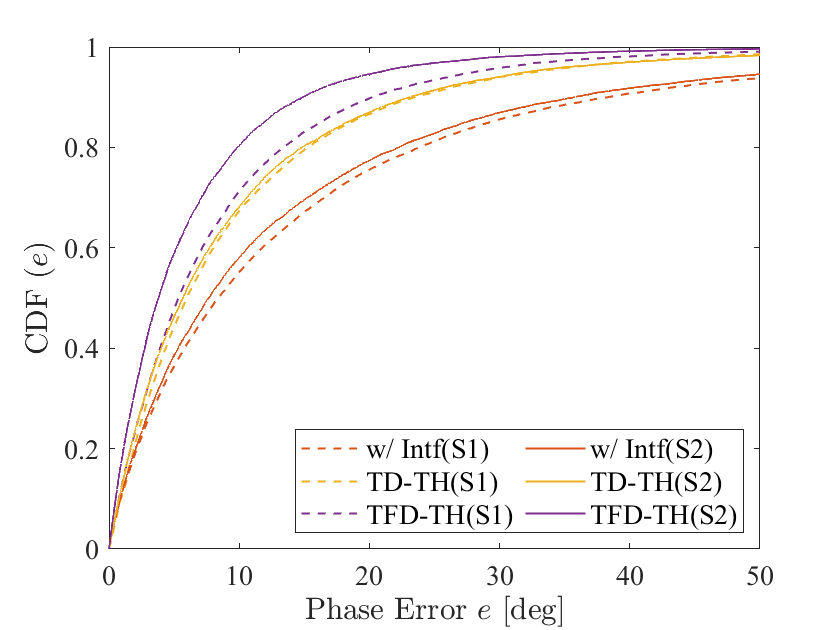}
\centerline{(c)}\medskip
\end{minipage}
\caption{Evaluation results of TD-TH and TFD-TH interference mitigation under interference scenario 1 (S1) and scenario 2 (S2): (a) PD, (b) SINR, and (c) phase error CDF.}
\label{fig:PD}
\end{figure*}

To evaluate the performance of \gls{tdth} and \gls{tfdth} methods in S1 and S2, the \gls{pd}, \gls{sinr}, and phase error \gls{cdf} are defined. Notably, some targets even cannot be detected by a typical detector under nominal conditions. Therefore, we refined the target cells in the \gls{rd} map to focus on only detectable targets, chosen based on a nominal condition detector threshold, e.g., the threshold leading to a probability of false alarm equal to $0.001$. The \gls{pd} is then derived from these detectable targets and the \gls{sinr} is derived by computing the average power of these targets against the noise power present in the \gls{rd} map. The phase of target bins in the \gls{rd} map plays a crucial role in subsequent target angle estimation. To measure the impact of interference mitigation on this, we defined the phase error \gls{cdf}: 
\begin{equation}
    \begin{split}
        \textrm{CDF}(e)= \textrm{Prob}\left\{ \left\lvert \angle\mathbf{Y}(i,j)-\angle\mathbf{Y}_c(i,j)  \right\rvert \leq e, \,\forall (i,j)\in \Theta  \right\}
    \end{split}
\end{equation}
Here, $\Theta$ represents the set of all target bins in the \gls{rd} map and the phase error $\left\lvert \angle\mathbf{Y}(i,j)-\angle\mathbf{Y}_c(i,j)  \right\rvert$ is scaled between $[0,180)$ degree.

In Fig.~\ref{fig:PD}, we present the impact of interference mitigation techniques, \gls{tdth} and \gls{tfdth}, on \gls{pd}, \gls{sinr}, and phase error \gls{cdf} as the probability of surrounding vehicles equipped with interfering radars, referred to as \textit{probability of radar interference}, increases. When the probability of radar interference reaches $100\%$, all vehicles within the scenario act as interference sources. The curves in Fig.~\ref{fig:PD} represent how the evaluation metrics change with different probability of radar interference. Each point on these curves represents the average result from $100$ Monte Carlo experiments. Comparing the curves with and without interference, the presence of interference leads to a notable decline in \gls{sinr} and a significant deterioration in the \gls{pd} of the victim radar. The target phase also experiences significant degradation due to interference. Upon employing interference mitigation techniques, both \gls{tdth} and \gls{tfdth} enhance \gls{sinr}. This leads to an improved \gls{pd} and reduced phase error. Yet, as we vary the probabilities of radar interference, \gls{tfdth} consistently outperforms \gls{tdth}. The superiority of \gls{tfdth} lies in its ability to preserve more target information within the time-frequency domain, as elaborated in section~\ref{sec:th}. \Gls{tfdth} shows superior performance in S2 compared to S1. This is due to the fact that, under uncorrelated interference conditions, the lower time resolution of the spectrogram makes TFD-TH's outcomes more aligned with those of \gls{tdth}. However, given the multiple interferences in S1, which falls into a kind of severe 3-interferer condition as illustrated in section~\ref{sec:th}, TFD-TH's performance remains superior to TD-TH.

\section{Conclusion}
In this paper, we have evaluated the performance of \gls{tdth} and \gls{tfdth} in dense traffic scenarios with three different types of interference. Our numerical results highlight that \gls{tfdth} consistently outperforms \gls{tdth} as the number of interference increases. This performance gap widens notably when the correlation between the interfering radar signal and the victim radar signal intensifies. The key to \gls{tfdth}'s superior performance lies in its ability to execute an accurate ``surgical removal" of interference signals within the time-frequency domain. Interestingly, the time-frequency representation utilized by \gls{tfdth} translates the signal into an image-like structure. This unique transformation unveils a promising avenue for leveraging data-driven techniques, such as deep learning, to remove interference in the time-frequency domain.

\newpage
\bibliographystyle{unsrt}
\bibliography{ref}
\end{document}